\DeclareUnicodeCharacter{FB01}{fi}
\documentclass[10pt, conference, letterpaper]{IEEEtran}

\usepackage{fancyhdr}
\usepackage{epsfig}
\usepackage{threeparttable}
\usepackage{epsf,epsfig}
\usepackage{amsthm}
\usepackage[cmex10]{amsmath}
\usepackage{amssymb, amsfonts}
\usepackage[noadjust]{cite}
\usepackage{dsfont}
\usepackage{subfigure}
\usepackage{color}
\usepackage{soul}
\usepackage{amssymb}
 \usepackage{accents}
  \usepackage{algorithm}
    \usepackage{url}
    \usepackage{framed} 
	\usepackage[colorlinks,
	linkcolor=black,
	anchorcolor=black,
	citecolor=black]{hyperref}

\newtheorem{remark}{\bf Remark}
\def\phi{\varphi}

\def\({\left(}
\def\){\right)}

\def\b0{{\mathbf{0}}}

\DeclareMathSizes{10}{9}{5}{3}

\usepackage{mathrsfs}
\usepackage{algorithm}
\usepackage{algorithmic}
\IEEEoverridecommandlockouts
\begin{document}	

\title{\huge Wireless Image Transmission with Semantic and Security Awareness\vspace{-5pt}}

% \author{\IEEEauthorblockN{Maojun Zhang\IEEEauthorrefmark{4},
% 	     Guangxu Zhu\IEEEauthorrefmark{1}, 
% 	     Shuai Wang\IEEEauthorrefmark{2}
% 	     Jiamo Jiang\IEEEauthorrefmark{5}
%               Caijun Zhong\IEEEauthorrefmark{4},
%               Shuguang Cui\IEEEauthorrefmark{3}\IEEEauthorrefmark{1}
%              }
% %             \vspace{-0cm}
% \IEEEauthorblockA{\IEEEauthorrefmark{4} College of information Science and Electronic Engineering, Zhejiang University, Hangzhou, China\\ 
% \IEEEauthorblockA{\IEEEauthorrefmark{1}Shenzhen Research Institute of Big Data, Shenzhen, China\\ 
% %Email: gxzhu@eee.hku.hk, huangkb@eee.hku.hk
% }
% \IEEEauthorblockA{\IEEEauthorrefmark{2} Southern University of Science and Technology, Shenzhen 518055, China}
% \IEEEauthorblockA{\IEEEauthorrefmark{5} China Academy of Information and Communications Technology, Beijing, China\\ 
% }
% \IEEEauthorblockA{\IEEEauthorrefmark{3}FNii and SSE, The Chinese University of Hong Kong (Shenzhen), Shenzhen, China\\ 
% 	%Email: gxzhu@eee.hku.hk, huangkb@eee.hku.hk
% }
% \IEEEauthorblockA{Email: zhmj@zju.edu.cn, gxzhu@sribd.cn, wangs3@sustech.edu.cn, \\ jiangjiamo@caict.ac.cn, caijunzhong@zju.edu.cn, shuguangcui@cuhk.edu.cn}
% }
%\vspace{-5mm}}

\author{Maojun Zhang, Yang Li, Zezhong Zhang, Guangxu Zhu, Caijun Zhong 
\thanks{M. Zhang and C. Zhong are with the College of information Science and Electronic Engineering, Zhejiang University, Hangzhou, China. (Email: zhmj@zju.edu.cn, caijunzhong@zju.edu.cn). 
Y. Li is with China Academy of Information and Communications Technology, Beijing, China (Email: liyang3@caict.ac.cn). 
Z. Zhang is with The Chinese University of Hong Kong (Shenzhen), Shenzhen, China (Email: zhangzezhong@cuhk.edu.cn). 
G. Zhu is with the Shenzhen Research Institute of Big Data, Shenzhen, China (Email: gxzhu@sribd.cn).
}}
\maketitle

%Update on Dec. 23:

\begin{abstract}
  Semantic communication is an increasingly popular framework for wireless image transmission due to its high communication efficiency. 
With the aid of the \emph{joint-source-and-channel} (JSC) encoder implemented by neural network, 
semantic communication directly maps original images into symbol sequences containing semantic information. 
%the original image is directly mapped into a symbol sequence containing semantic information. 
Compared with the traditional separate source and channel coding design used in bit-level communication systems, 
semantic communication systems are known to be more efficient and accurate especially in the low signal-to-the-noise ratio (SNR) regime. 
This thus prompts a critical while yet to be tackled issue of security in semantic communication: 
it makes the eavesdropper much easier to crack the semantic information as it can be retrieved even in a highly noisy channel.  
%current modular design, 
%the semantic transmission efficiency can be dramatically improved. % compared with the traditional modular transmission scheme. 
%The current objective of semantic communication is to minimize the transmission distortion at destination user (Bob). 
%However, privacy protection, which is one of the main objective in communication design, has not been studied well yet.  
In this letter, 
we develop a semantic communication framework 
that accounts for both semantic meaning decoding efficiency and its risk of privacy leakage. 
%that satisfies the both aforementioned objectives. 
To this end, targeting wireless image transmission, 
we propose an JSC autoencoder featuring residual structure for efficient semantic meaning extraction and transmission, 
%image transmission. 
and the training of which is guided by a well-designed loss function
%and on the other hand, propose 
%to resist the eavesdropper (Eve), we propose 
%a data-driven scheme 
that can flexibly regulate the efficiency-privacy trade-off. 
Extensive experimental results are provided to show the effectiveness and robustness of the proposed scheme.
  %Semantic communication is a promising framework for wireless image transmission. The main gain is that it can still achieve satisfactory performance even when the wireless environment is poor. However, semantic communication will bring another problem. Due to the strong robustness, the eavesdropper, who is usually in a scenario of low signal-to-noise ratio (SNR), can decode the information at the semantic level as well, which brings serious privacy problems. In this letter, we on the one hand propose an efficient joint source and channel coding (JSCC) scheme for semantic image transmission. On the other hand, to resist the eavesdroppers, we propose two effective schemes to strengthen privacy protection. The experimental results shows the effectiveness of the proposed scheme. 
  %This letter reveals the privacy issues in the current semantic communication system. 
\end{abstract}

\vspace{-2mm}
\section{introduction}
%The rapid development of artificial intelligence (AI) prompts a new direction for the future evolution of wireless networks \cite{gunduz2022beyond}. 
The wide success of artificial intelligence (AI) in every perspectives of our society has also driven the rapid advancement in wireless communications \cite{gunduz2022beyond}. 
Recently, 
as a consequence of the fusion of AI and communication, 
%through the deep combination of AI and communication, 
a novel paradigm, called semantic communication, has received great attention. 
%Inherited from 
Building on the deep learning based end-to-end communication system \cite{ye2020deep}, 
semantic communication further introduces efficient semantic encoder network, 
so that the essential semantic information instead of the raw data can be extracted, 
encoded and delivered to the receiver, which is believed to be a more efficient and effective way to convey information in the next generation wireless networks \cite{xie2021deep}. 
%and then transmits the important semantic information instead of the original data to  the receiver, thus realizing the information transmission at the semantic level \cite{xie2021deep}. 

%Specifically, 
In particular, semantic communication has shown promising gain in image transmission task. 
In classic bit-level communication, the images are first compressed into binary sequences by source coding algorithms (e.g., JPEG, JPEG2000, BPG), followed by channel coding schemes (e.g., Turbo, LDPC) that add certain redundancy to combat against the random channel perturbation, 
and after that the codewords are modulated into symbol sequences for reliable transmission. 
Such a separate source and channel coding scheme is hard to guarantee the optimality of the whole system in terms of rate-distortion trade-off. 
Prompted by this,  authors in \cite{bourtsoulatze2019deep} first proposed a \emph{joint-source-and-channel-coding} (JSCC) method for wireless image transmission, where the images were directly mapped into complex-valued symbols through a well-trained neural network. 
To further improve the quality of reconstructed image, the feedback and multi-layer bandwidth-agile design were subsequently proposed in \cite{kurka2020deepjscc,kurka2021bandwidth}. 
In addition, since semantic communication aims to deliver semantic meaning instead of perfectly reconstructing the original source at the receiver. 
%is on the semantic reconstruction, 
For image transmission tasks, the commonly-used pixel similarity (e.g., PSNR \cite{bourtsoulatze2019deep}) is  no longer appropriate to describe the goodness of semantic communication. 
Given this, some new reconstruction 
performance metric customized for semantic communication 
%criteria based on semantic perceptual 
were proposed in \cite{wang2022perceptual,huang2022towards},  
where semantic communication exhibits remarkably higher efficiency and accuracy than the bit-level communication, especially in the low signal-to-the-noise ratio (SNR) regime. 

Like every coin has two sides, accompanying the good performance in the low SNR regime is the higher risk of privacy leakage, 
as it implies that the eavesdroppers can crack the semantic information more easier even through a highly noisy channel. 
This thus prompts a critical issue regarding secure semantic communications. 
%Despite its great potential, semantic communication still faces several challenges in its practical deployment. 
%Although semantic communication has shown its great potential for future intelligent communication scenarios, there are still some key concerns about its practical deployment. 
% The semantic communication system consists of neural networks, which has strong dependence on training data. 
% Thus the efficient training algorithm should be developed \cite{zhang2022deep}. 
% %Therefore, the additional adaptive design is required to cope with both new data and new scenarios \cite{zhang2022deep}. 
% In addition, secure transmission has always been one of the pursuits in the communication system, 
% however, {\emph{whether semantic communication is secure}} and \emph{how to achieve secure semantic communication} have not been studied well yet. 
%until now the privacy related issues have not be well studied. 
%Finally and most importantly, whether semantic communication is secure and how to achieve secure semantic communication is still a blank field. 
%For example, security has always been hot topics in wireless communications due to its inherent broadcast nature, but not yet been well understood in the context of semantic communication. 
To design security-aware semantic communication systems, 
%in addition to maximizing 
one needs to balance the trade-off between 
the transmission efficiency at the destination user (Bob), 
and the information leakage to the eavesdropper (Eve). 
In classic bit-level communication systems, the secure channel capacity, rather than channel capacity, serves as the main performance metric of interest  to ensure security. %in secure communications. 
The theoretical analysis of secure capacity was presented in \cite{gopala2008secrecy,liang2008secure} targeting bit-level. 
Building on it, the secrecy capacity region can be derived, and secure transmission can be achieved by proper transmission power control and specific channel coding designs \cite{besser2019wiretap,fritschek2019deep,tung2022deep}. 
Nevertheless, in semantic communication, the ``black-box'' nature of JSCC block implemented by neural networks makes the derivation of secure channel untractable if not impossible. 
Therefore, the existing secure communication schemes
% based on existing 
building for bit-level
communication systems cannot be directly applied to semantic communication systems,  
leaving the secure semantic communication remains a largely uncharted area. 

As discussed above, there are two basic objectives in secure semantic communication systems, namely, 
the one concerns efficiency, i.e., the semantic recovery quality at Bob; 
and the other concerns privacy, i.e., the semantic leakage to Eve. 
This gives rise to a fundamental trade-off between efficiency and privacy. 
In this letter, we develope a secure semantic communication framework that 
accounts for both objectives above. 
%satisfies the above objectives both. 
Firstly, we propose an efficient \emph{joint-source-channel} (JSC) autoencoder 
featuring the cascading of residual block with convolution layer for efficient semantic meaning extraction and transmission, 
and the training of which is guided by a well-designed loss function that can flexibly balance the efficiency-privacy trade-off. 
%for semantic encoding.  
%that can efficiently extract semantic information. 
%With the powerful 
%Featuring the residual structure, 
% Featuring the alternating structure of residual block and convolution layer, 
% the proposed JSC autoencoder can extract the semantic information efficiently. 
% Secondly, we show that the privacy leakage issue does exist in current semantic communication system and even worse than that in the digital communication system as the JSCC design has strong robustness to channel quality. 
% Inspired by prior work in secure channel capacity \cite{gopala2008secrecy,liang2008secure}, we propose a data-driven scheme that balances the said efficiency-privacy. %trade-off between the reconstruction quality and the privacy leakage. 
%solve the privacy problem from a data-driven perspective, 
%propose a training objective from a data-driven perspective. 
%Experiments 
Extensive experiments are conducted to
show that the proposed JSCC scheme can significantly outperform the traditional separate source and channel coding scheme in the low SNR regime, in the meanwhile, prevent privacy leakage at the semantic level, thus achieving the desired efficient and secure semantic communication.
\vspace{-2mm}
\section{System Model}\label{sec: system model}
In this section, we present the downlink semantic communication system for image transmission, 
and put forth the privacy issue caused by Eve. 
\subsection{Semantic Transmitter}
\begin{figure}[h]
  \centering
  \includegraphics[width=1\linewidth]{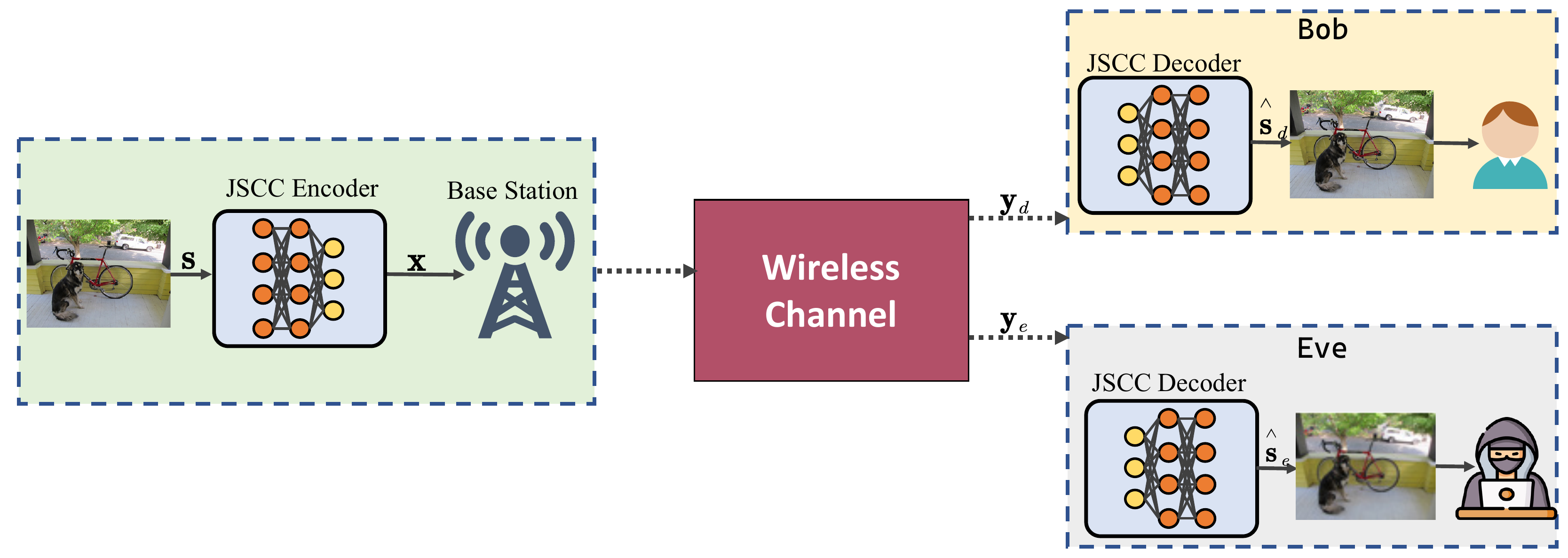}
  \caption{Illustration of the semantic communication system for image transmission}\label{fig:semantic communication system}
  %\vspace{-10mm}
  \label{fig:semantic communication system}
  \end{figure}
As shown in Fig \ref{fig:semantic communication system}, 
the \emph{base station} (BS) to conﬁdentially transmit the image $\mathbf{s}$ to the legitimate user (Bob), in the presence of a passive eavesdropper (Eve).  
%The communication is overheard through a second channel by an eavesdropper (Eve). 
Different from the conventional  separate source coding (e.g., JPEG, BPG) and channel coding (e.g., Turbo, LDPC code) design, 
the compression and anti-jamming are implemented by the \emph{joint-source-and-channel} (JSC) encoder composed of deep neural networks (DNNs). 
The encoding process is given as follows: 
\begin{align}\label{eq: semantic encoder}
  \mathbf{x} = f\left(\mathbf{s};\boldsymbol{\theta}\right), 
\end{align}
where $\boldsymbol{\theta}$ and $\mathbf{x}\in \mathbb{R}^{M\times 1}$ denote the trainable parameters in JSC encoder and the latent semantic representation of the image source $\mathbf{s}$, respectively. 

Considering the transmit power limitation, we have the following power constraint on the transmitted signal, i.e., $\frac{1}{M}\mathbb{E}\left\{x_i^2\right\}\leq p$. 
%\begin{align}\label{eq: limitation on the latent representation}
  %$\left|x_i\right|^2\leq p$,  
  %-\sqrt{p} \leq\Re\left(\left[\mathbf{x}\right]_i\right),\Im\left(\left[\mathbf{x}\right]_i\right)\leq \sqrt{p},~\forall i, 
%\end{align}
%where %$\Re\left(x\right)$ and $\Im\left(x\right)$ denotes the real part and the imaginary part of element $x$, 
% $p$ denotes the maximum transmit power of each element.  
%\vspace{-4mm}
\subsection{Wireless Transmission}\label{subsec: wireless transmission}
%\vspace{-2mm}
% We consider two channel models. 
% The first is the single antenna \emph{additive white Gaussian noise} (AWGN) channel, which is widely used in prior works on secure communication, where the noise level of the wiretap channel is assumed to be higher than the legitimate channel from BS to Bob. 
% %\emph{additive white Gaussian noise} (AWGN) system is widely discussed in the former works on secure communication, in which noise level of the wiretap channel to Eve is set higher than the channel from BS to Bob. 
% In addition, we also consider a more practical single user \emph{multiple-input-single-output} (MISO) antenna system, 
% where precoding techniques can be exploited to relax the noise power requirement in AWGN system. 
%in which the transmission heterogeneity of Bob and Eve is generated by precoding.
% \subsubsection{AWGN}
We consider the \emph{additive white Gaussian noise} (AWGN) channel. 
%We consider the Gaussian wiretap channel model. 
The received signal of Bob through the legitimate AWGN channel is given by 
\begin{align}\label{eq:}
  \mathbf{y}_b = \mathbf{x} + \mathbf{n}_b,
\end{align}
where $\mathbf{n}_b\sim \mathcal{N}\left(0,\sigma_b^2\mathbf{I}\right)$, $\sigma_b^2$ is the average noise power. 

Similarly, Eve can receive the information through the wiretap channel as follows:  
%We call them potential eavesdropper, 
%The received signal of Eve is given by  
\begin{align}
  \mathbf{y}_e = \mathbf{x} + \mathbf{n}_e,
\end{align}
where $\mathbf{n}_e\sim \mathcal{N}\left(0,\sigma_e^2\mathbf{I}\right)$, $\sigma_e^2$ is the average noise power. 
Generally, as in \cite{liang2008secure,besser2019wiretap}, we assume  that the wiretap channel between BS and Eve is worse than the channel between BS and Bob, i.e., $P = \frac{\sigma_e^2}{\sigma_b^2}\gg 1$. 
\subsection{Semantic Receiver}\label{subsec: semantic receiver}
%\subsection{Compression Model}
%With the received signals, 
In the receiver side, 
both Bob and Eve can try to decode the image as follows: %corresponding one using the received signals, which is given as follows. 
\begin{align}
  \widehat{\mathbf{s}}_b = g\left(\mathbf{y}_b;\boldsymbol{\Theta}_b\right),\qquad \widehat{\mathbf{s}}_e = g\left(\mathbf{y}_e;\boldsymbol{\Theta}_e\right). 
\end{align} 
{We note that both JSC encoder and decoder can only be deployed after sufficient training, while the unbearable communication overhead will be introduced as a cost.  
Moreover, there is a strong demand for serving multiple users in semantic communication system. Given these, 
sharing the JSC decoder publicly should be a proper solution for alleviating the training burden, i.e., $\boldsymbol{\Theta}_e =\boldsymbol{\Theta}_d $, as users in the cell can collaborate to improve the JSC decoder through federated learning.  
%we adopt that the JSC decoder is shared publicly in the cell. 
However, the shared JSC decoder raises critical privacy issue that 
it makes Eve easy to crack the semantic information  
as it can be retrieved even in a quite noisy channel. 
%Eve is likely to decode the image even when the wiretap channel is weak.  
}%In this way, the training problem can be solved in an distributed manner, i.e., federated learning \cite{shi2021semantic}. 
%It is assumed that the JSC decoder is available to both Bob and Eve, which raises critical privacy issue, that Eve is likely to decode  the original image even when the wiretap channel is weak. 
We shall tackle such issues in the subsequent section. 
%Responding to this, we will address it through custom training strategy. 
% Note that, the privacy issue can be well addressed by aligning different JSC decoder among devices, i,e, $\boldsymbol{\Theta}_d\neq \boldsymbol{\Theta}_e$. 
% In this way, the eavesdropper with $\boldsymbol{\Theta}_e$ cannot recover $\mathbf{s}$ from $\mathbf{y}_e$. 
% However, 
% Here we should first argue that the model should be shared across devices, thus the 

%We adopt an unbiased stochastic sparsification scheme \cite{wangni2018gradient}. 
\section{Proposed Method}
In this section, 
we first propose a JSC autoencoder 
%featuring the alternating structure of residual block and convolution layer 
featuring the cascading of residual block with convolution layer 
to extract the semantic information efficiently. 
Given the potential privacy leakage, we then propose a data-driven scheme that balances the efficiency-privacy trade-off. %the trade-off between the reconstruction quality of Bob and the privacy leakage to Eve. 
%we will discuss the two basic problem in semantic communication system described in Section \ref{sec: system model}. 
%On the one hand, a novel JSC encoder and decoder is proposed to extract the semantic information efficiently, 
%on the other hand, realizing that the issue of privacy leakage from the robustness of semantic communication system in low SNR, we propose two ways to preserve privacy while still guarantee the transmission quality of destination user. 
%\textbf{Efficient Semantic Extraction} and \textbf{Privacy Preserving}
\subsection{JSC Autoencoder Design}
\vspace{-3mm}
\begin{figure}[h]
  \centering
  \includegraphics[width=1\linewidth]{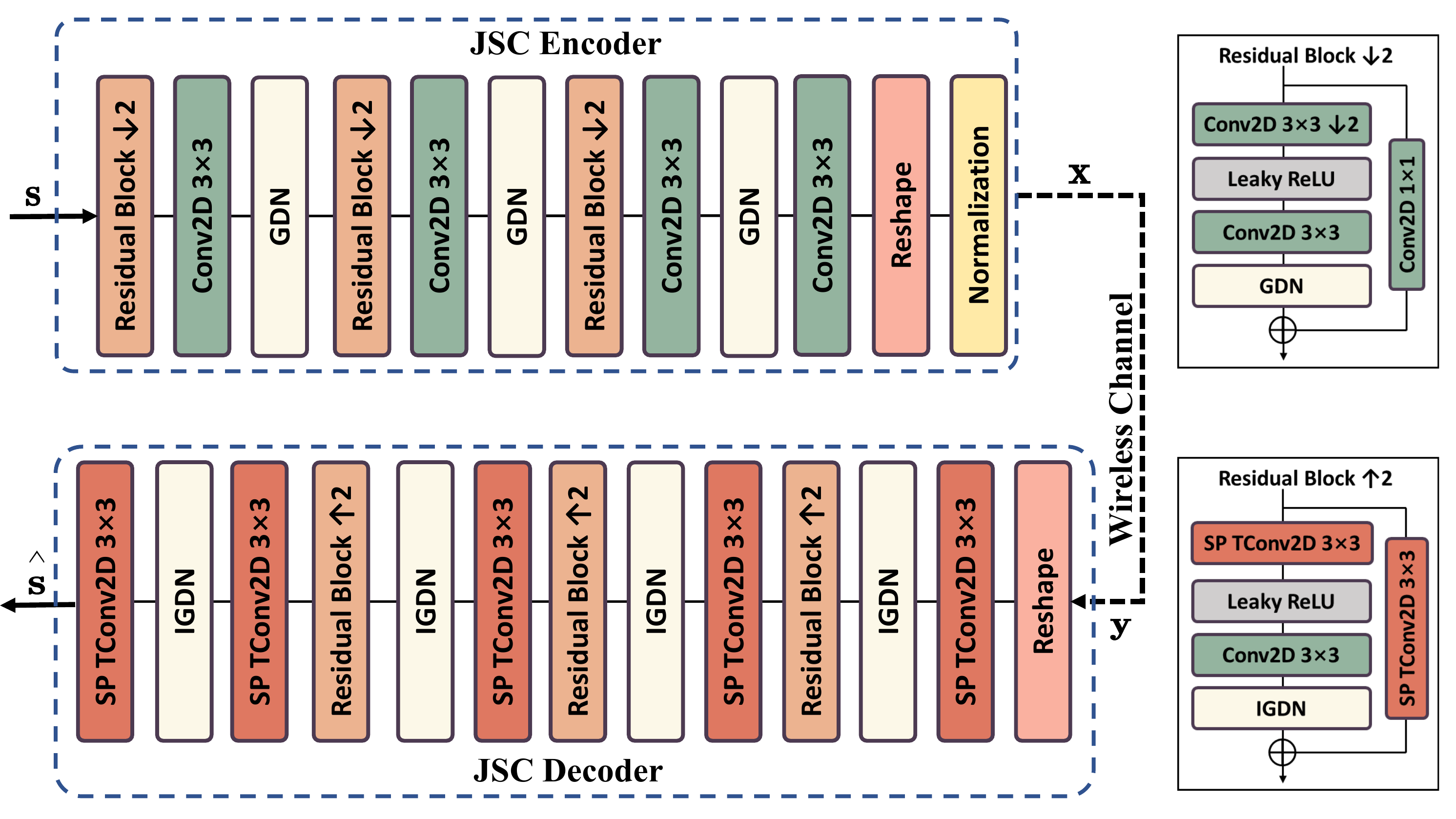}
  \caption{Network architecture of the JSC autoencoder}\label{fig:network structure}
  \vspace{-1mm}
  \label{Fig:1}
  \end{figure}
  \vspace{-3mm}
% As shown in Fig. \ref{fig:network structure},  
% we adopt the network proposed in \cite{cheng2020learned}. 
% In the encoder side, the x(the network description) in this paragraph. 
The network architecture of JSC autoencoder is shown in Fig. \ref{fig:network structure}. 
%Followed by \cite{cheng2019deep}, compared with the previous work that adopts the full convolutional structure \cite{bourtsoulatze2019deep,kurka2020deepjscc,wang2022perceptual,kurka2021bandwidth}, 
As in \cite{cheng2019deep}, 
the residual blocks are added to improve the model performance and training stability. 
In the encoder part, we adopt the method of alternately cascading the residual block with convolution layer, and downsampling the input image three times through residual block. 
In addition, all the intermediate results are normalized by generalized normalization transformations (GDN) \cite{balle2015density}, which is widely used in image compression. 
The network structure of the decoder is similar to the encoder, while the sub-pixel convolution layer \cite{shi2016real} is adopted to reconstruct the image. 
Compared with the transposition convolution layer, it can improve the resolution of the obtained image through learning, thus improving the reconstruction performance. 

Note that, unlike in traditional communication systems where perfect recovery of $\mathbf{x}$ from $\mathbf{y}_b$ is pursued, we train the autoencoders in an end-to-end way,  
%To improve restoration quality of the transmitted pictures, both the JSC encoder and the JSC decoder should be trained in an end-to-end manner, 
as such, the image compression and channel adaption can be achieved by using the following loss function,  
\begin{align}\label{eq:mse loss function}
  \mathcal{L}_1=\frac{1}{B}\sum_{i=1}^B d\left(\mathbf{s}_i,\widehat{\mathbf{s}}_i\right),
  %\min_{\boldsymbol{\theta},\boldsymbol{\Theta}} \left\|\mathbf{s}-\widehat{\mathbf{s}}\right\|_2^2. 
\end{align}
where $B$ denotes the batch size, $d\left(\mathbf{s}_i,\widehat{\mathbf{s}}_i\right)=\left\|\mathbf{s}_i-\widehat{\mathbf{s}}_i\right\|^2$ is the \emph{mean squared-error} (MSE) distortion between the reconstructed image and the raw image. 
%Using the dataset that is both owned by the BS and users, the optimal $\boldsymbol{\theta}^*$ and $\boldsymbol{\Theta}^*$ can be attained by minimizing the empirical distortion. %over the public dataset. 
%For the design of JSC autoencoder, 
%where the objective is to minimize the MSE distortion between $\widehat{\mathbf{s}}_i$ and $\mathbf{s}_i$. 
%This shows the difference between the semantic communication and the traditional wireless communication. 
\subsection{Privacy-aware Design}
\begin{figure}[htb!]
    \centering
    \subfigure[Original image]{
      \begin{minipage}[t]{0.3\linewidth}
        \centering
        \includegraphics[width=1\linewidth]{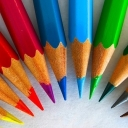}\\
        \vspace{0.02cm}
        % \includegraphics[width=1\linewidth]{Figures/Base_AWGN/d2dSNR_10/origin_9.png}\\
        % \vspace{0.02cm}
      \end{minipage}%
    }%
    \subfigure[BPG-Turbo-64QAM]{
      \begin{minipage}[t]{0.30\linewidth}
        \centering
        \includegraphics[width=1\linewidth]{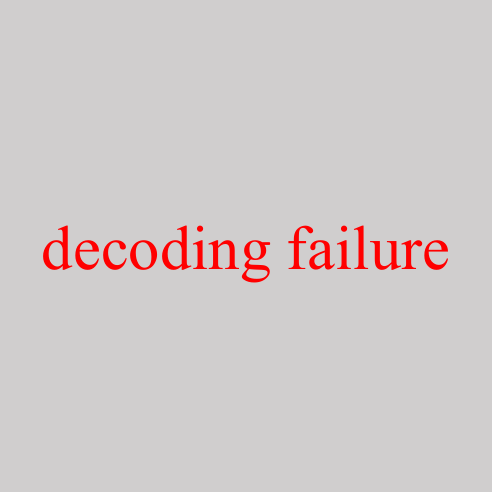}\\
        \vspace{0.02cm}
        % \includegraphics[width=1\linewidth]{Figures/Base_AWGN/d2eavSNR_10/predict_9.png}\\
        % \vspace{0.02cm}
      \end{minipage}%
    }%
    \subfigure[JSCC with (\ref{eq:mse loss function})]{
      \begin{minipage}[t]{0.3\linewidth}
        \centering
        \includegraphics[width=1\linewidth]{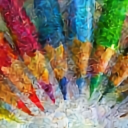}\\
        \vspace{0.02cm}
        % \includegraphics[width=1\linewidth]{Figures/Base_AWGN/d2dSNR_10/predict_9.png}\\
        % \vspace{0.02cm}
      \end{minipage}%
    }%
    \centering
    \caption{The reconstructed image produced by Eve (${\rm SNR}_{\rm Eve}=0{\rm dB}$)}\label{fig: privacy issue}
    \vspace{-0.2cm}
\end{figure}
% As we have discussed in Section \ref{subsec: semantic receiver}, when the BS side wants to transmit images to the destination user, 
% the non-destination user also called potential eavesdroppers can partially decode the image as well, causing privacy leakage. 
% In this subsection, we propose two ways to address this problem. 
% As former works has revealed  \cite{xie2021deep,bourtsoulatze2019deep}, the semantic communication system could achieve satisfactory performance in poor wireless environment, i.e., low transmission SNR. 
% Moreover, as the deep learning-based end-to-end communication maps the original information to a continuous transmission symbol, traditional binary sequence-based encryption is no longer applicable. 
% This brings the critical privacy issue, that is, Eve can partially decode the original source with the powerful JSC decoder even when the channel is much more poor than Eve, as shown in Fig. \ref{fig:example of AWGN} (c) and Fig \ref{fig:example of MISO} (c).  
% This reminds us to take the privacy leakage into account in the design of semantic communication system. 
% The overall privacy preserving objective is given by
As reported in \cite{xie2021deep,bourtsoulatze2019deep}, one of the advantages of semantic communication is that the satisfactory performance can be achieved even in the low SNR regime. 
As shown in Fig. \ref{fig: privacy issue}\footnote{The detailed settings are given in Section \ref{sec: simulation setting}.}, %the \emph{signal-to-noise ratio} (SNR) of the wiretap channel is set to 10dB,  
with the poor wiretap channel, Eve from the conventional image transmission system cannot decode anything, while with the powerful JSC decoder, the Eve from the semantic communication system can still crack the semantic information. 
It shows that the semantic communication system does have higher risk of privacy leakage than bit-level communication. %more serious privacy issue. 
%In the context of secure image transmission, it makes easier for Eve to recover the image, as shown in Fig. \ref{fig:example of AWGN}(c) and Fig \ref{fig:example of MISO}(c). 

% As former works has revealed  \cite{xie2021deep,bourtsoulatze2019deep}, the semantic communication system could achieve satisfactory performance in poor wireless environment, i.e., low transmission SNR. 
% Once the JSC decoder is available in the Eve side, Eve can partially decode the original source through the received signal from the wiretap channel, as shown in Fig. \ref{fig:example of AWGN} (c) and Fig \ref{fig:example of MISO} (c).  
To address this issue, 
similar to secure channel capacity, 
an intuitive way is to take the reconstruction quality of Eve into account of the training objective.   
The loss function with privacy aware is given by 
\begin{align}\label{eq: secure mse loss definition}
  \mathcal{L}_2 = \frac{1}{B}\sum_{i=1}^B \left[d\left(\mathbf{s}_i,\widehat{\mathbf{s}}_{d,i}\right)- \lambda\cdot d' \left(\mathbf{s}_i,\widehat{\mathbf{s}}_{e,i}\right)\right],
\end{align}
where $\lambda$ is the weighting factor, $d'(\cdot)$ characterizes the privacy leakage to Eve. 

Then, the main challenge is to give a proper design of $d’(\cdot)$. 
There are two principles for it. Firstly, $d’(\cdot)$ does not have to be the same form with $d$, as privacy information may have various definitions. 
Secondly, there exists a trade-off between the reconstruction quality of Bob and privacy leakage to Eve.  
We should minimize the reconstruction distortion of Bob while protecting privacy to a certain degree. 
Considering these, we propose the following criterion of privacy leakage, 
\begin{align}\label{eq:privacy aware design}
  d'\left(\mathbf{s}_i,\widehat{\mathbf{s}}_{e,i}\right) = \left\{\begin{array}{lr}
    -d\left(\mathbf{0},\widehat{\mathbf{s}}_{e,i}\right)~~~~ d\left(\mathbf{0},\widehat{\mathbf{s}}_{e,i}\right)>\epsilon\\
    0~~ \qquad\qquad{\rm otherwise.}
  \end{array}\right.
\end{align}
where $\epsilon$ is the predefined indicator of privacy protection, $\mathbf{0}$ denotes the all black image with a same shape of $\mathbf{s}$. 
\begin{remark}
\emph{  Generally, the degree of privacy leakage can be characterized by the mutual information ${\rm I}(\mathbf{s}_{i};\widehat{\mathbf{s}}_{e,i})={\rm H}(\widehat{\mathbf{s}}_{e,i})-{\rm H}(\widehat{\mathbf{s}}_{e,i}|\mathbf{s}_{i})$. 
  However, ${\rm I}(\mathbf{s}_{i};\widehat{\mathbf{s}}_{e,i})$ is not tractable. 
  We instead minimize ${\rm H}(\widehat{\mathbf{s}}_{e,i})$, the upper bound of ${\rm I}(\mathbf{s}_i;\widehat{\mathbf{s}}_{e,i})$. 
  It can be achieved by forcing the image decoded by Eve to converge to the constant one, 
  i.e., the all-black image. 
  In addition, $d’\left(\cdot\right)$ serves as a penalty function for the training objective, for which we set a threshold, and if the privacy leakage exceeds $\epsilon$, $d’\left(\cdot\right)$ will corrects the training direction for privacy protection. 
}  % \emph{ Essentially, (\ref{eq:loss function of secure mse}) tries to balance two competing objectives.  
  %   On the one hand, the JSC autoencoder should try to reduce the transmission distortion on Bob, while on the other hand, 
  % \rv{  the JSC autoencoder should also try to make the image decoded by Eve as close to all black pictures as possible to prevent privacy leakage.     
  % }
\end{remark}

Combining (\ref{eq: secure mse loss definition}) with (\ref{eq:privacy aware design}), the novel loss function with privacy aware is presented, referred to as  \emph{secure mean squared-error} (SecureMSE). 
\section{Simulation Results}\label{sec: simulation setting}
In this section, we conduct a set of experiments evaluating the performance of the proposed scheme, 
including the reconstruction quality at Bob and the privacy leakage to Eve. 
The AWGN system  in Section \ref{subsec: wireless transmission} is first considered, where $P=\frac{\sigma_e^2}{\sigma_b^2}$ is set to 15dB. 
In addition, a more practical \emph{multiple-input-single-output} (MISO) system is also considered, where precoding techniques can be exploited to relax the noise power requirement in AWGN system. 
%We consider two communication systems as shown in Section \ref{subsec: wireless transmission}, e.g., AWGN and MISO channels at different channel SNR settings. 
%In AWGN system, we assume that $P=\frac{\sigma_e^2}{\sigma_b^2}=15{\rm dB}$. 
%In MISO system, each element of $\mathbf{h}_b$ and $\mathbf{h}_{e}$ is assumed to follow the standard Gaussian distribution, and $P=\frac{\sigma_e^2}{\sigma_b^2}=0{\rm dB}$, $N_t=8$. 
Specifically, $\mathbf{x}$ is normalized to $\sqrt{Mp}\mathbf{x}/\left\|\mathbf{x}\right\|_2^2$ to satisfy the average power constraints, with $p=1$. 
In all the presented figures, we denote ${\rm SNR}$ as the transmission signal-to-noise ratio at Bob side. 
%The ${\rm SNR}$ below denotes the transmission signal-to-noise ratio of Bob. 

\textbf{Dataset:} We use the Linnaeus 5 dataset for training and testing.\footnote{chaladze.com/l5}  The images have dimension $128\times 128 \times 3$. The whole image dataset is composed of 5 classes, including berry, bird, dog, flower, and other. There are 1200 training images, 400 testing images per class. 

\textbf{Performance Metric:} To measure the performance of the proposed scheme and the baseline schemes, we use the structural similarity index measure (SSIM) as the performance metrics, which is given below. 
\begin{align}\label{eq: SSIM}
  {\rm SSIM}(\mathbf{s},\widehat{\mathbf{s}})=\frac{\left(2\mu_{\mathbf{s}}\mu_{\widehat{\mathbf{s}}}+c_1\right)\left(2\sigma_{\mathbf{s}\widehat{\mathbf{s}}}+c_2\right)}{\left(\mu_{\mathbf{s}}^2+\mu_{\widehat{\mathbf{s}}}^2+c_1\right)\left(\sigma_{\mathbf{s}}^2+\sigma_{\widehat{\mathbf{s}}}^2+c_2\right)}
\end{align}
where $\mu_{\mathbf{s}}$, $\sigma_{\mathbf{s}}^2$, $\sigma_{\mathbf{s}\widehat{\mathbf{s}}}^2$ are the mean and variance of $\mathbf{s}$, and the covariance between $\mathbf{s}$ and $\widehat{\mathbf{s}}$, respectively. $c_1$ and $c_2$ are constants for numeric stability. 

\textbf{Training Setting:}  %The channels of each block in Fig. are given in Table I. 
We adopt the Adam optimizer, with the learning rate of 0.0001. 
the pretrained model is first obtained by using the ImageNet dataset with the loss function of MSE and the assumption of ideal transmission (i.e., the receiver can obtain $\mathbf{x}$ without noise.). 
Then the final model is obtained through training under specific channel and loss setting. 
All the number of filters in residual blocks and convolution layers are $128$.
The experiments are implemented by PyTorch and Python 3.8 on a Linux server with 2 NVIDIA RTX 3090 GPUs. 
\subsection{Reconstruction Evaluation}
\vspace{-1mm}
\begin{figure}[htb!]
  \centering
  \includegraphics[width=1\linewidth]{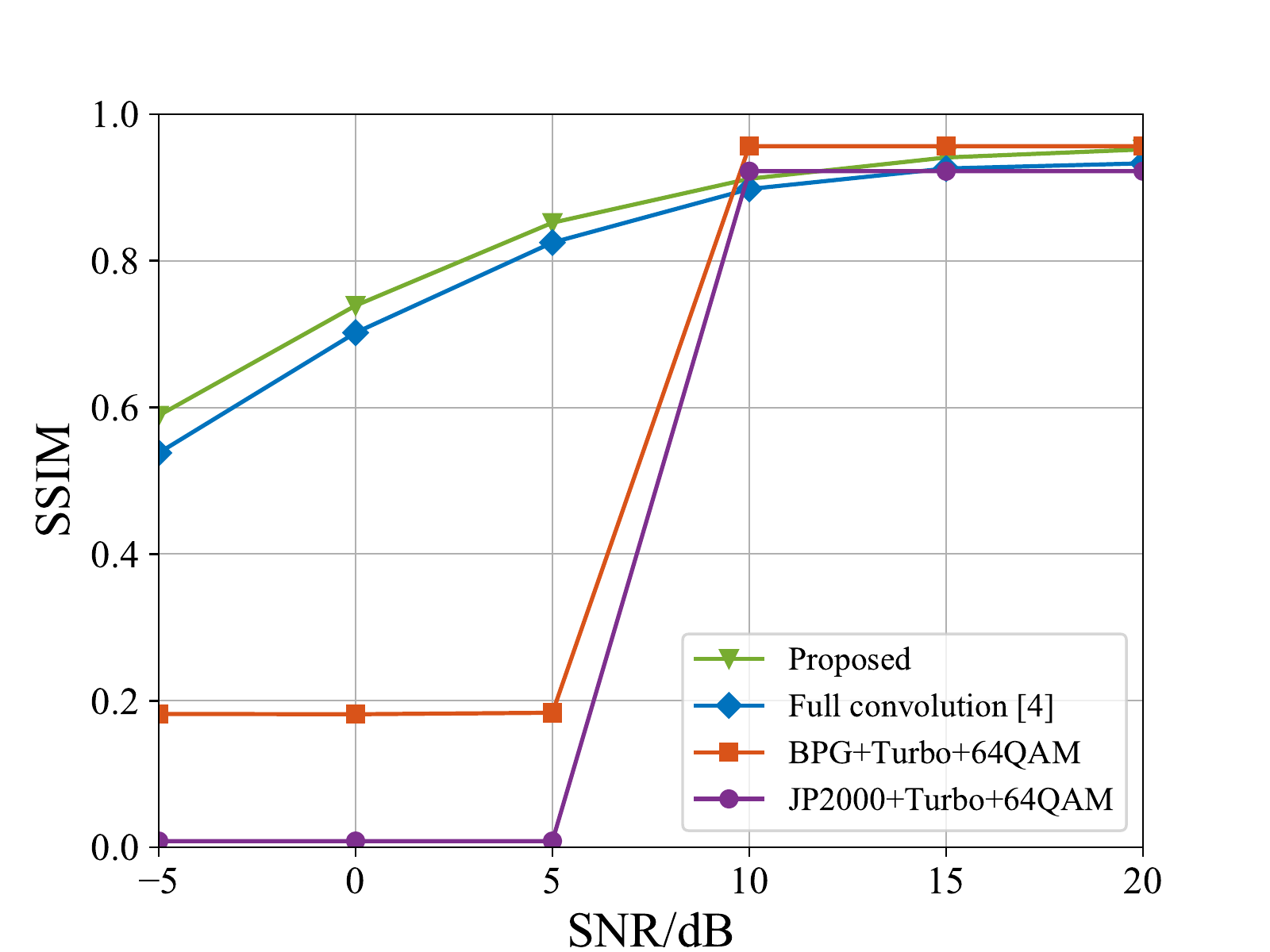}
  \caption{Performance comparison among different transmission schemes}\label{fig:base performance}
  \label{Fig:1}
  \end{figure}
In this subsection, we compare the reconstruction quality of the proposed schemes, the JSCC scheme in \cite{bourtsoulatze2019deep}, and the conventional schemes with separate source and channel coding. 
For the conventional schemes, two source coding schemes including JPEG2000 and BPG are considered.  
% For the source coding, we consider two well-known image compression algorithms, JPEG2000 and BPG. 
% The processed bit sequences are further processed by the Turbo codes, which is used to combat against the channel noise. 
% The bit sequence is finally modulated through 64 quadrature amplitude modulation (QAM). 
To ensure fairness, the same number of transmitted symbols are guaranteed. 
The results are presented in Fig. \ref{fig:base performance}.
It can be seen that for the two traditional schemes, the reconstruction quality is bad when the channel quality is poor (i.e., ${\rm SNR}\leq5{\rm dB}$). 
This is because the traditional scheme needs to represent the original picture as bit sequences. 
The poor channel leads to high bit error rate. 
For the compression and decompression schemes of BPG and JPEG2000 standards, the accumulation of bit errors will cause decoding failure. % which makes the conventional schemes fail to work. %directly lead to non decoding, thus the conventional schemes does not work. 
As for the JSC coding scheme, it transmits the most important semantic information in the form of symbols. 
Although there exist symbol errors, only semantic offset occurs. 
It has a relatively satisfactory performance under low SNR regime and maintains similar performance with traditional schemes under high SNR regime. 
Moreover, the alternating residual and convolutional structure outperforms the full convolution structure, which verifies the effectiveness of the proposed scheme. 

% Moreover, with the data-driven anti-interference characteristics of the neural network, it still has a relatively satisfactory performance under low SNR. 
% At the same time, under high SNR, the proposed scheme has similar performance with traditional schemes, which verifies the superiority of semantic communication.
 
\subsection{Security Evaluation}
\vspace{-5mm}
\begin{figure}[h]
  \centering
  \includegraphics[width=9cm]{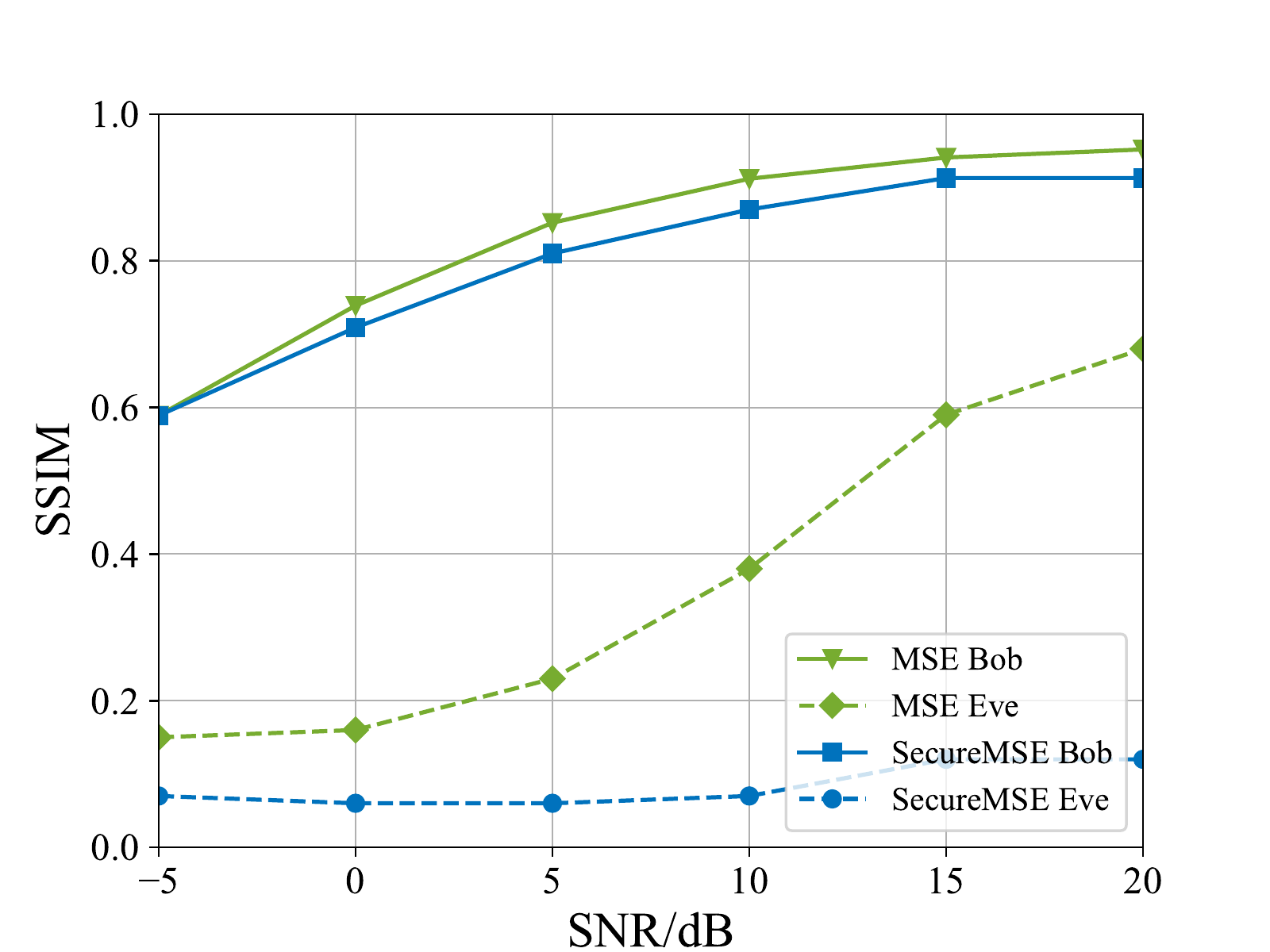}
  \caption{Performance comparison of Bob and Eve in AWGN system}\label{fig: secure SSIM AWGN}
  %\vspace{-10mm}
  \end{figure}
  \begin{figure}[!]
    \centering
    \includegraphics[width=9cm]{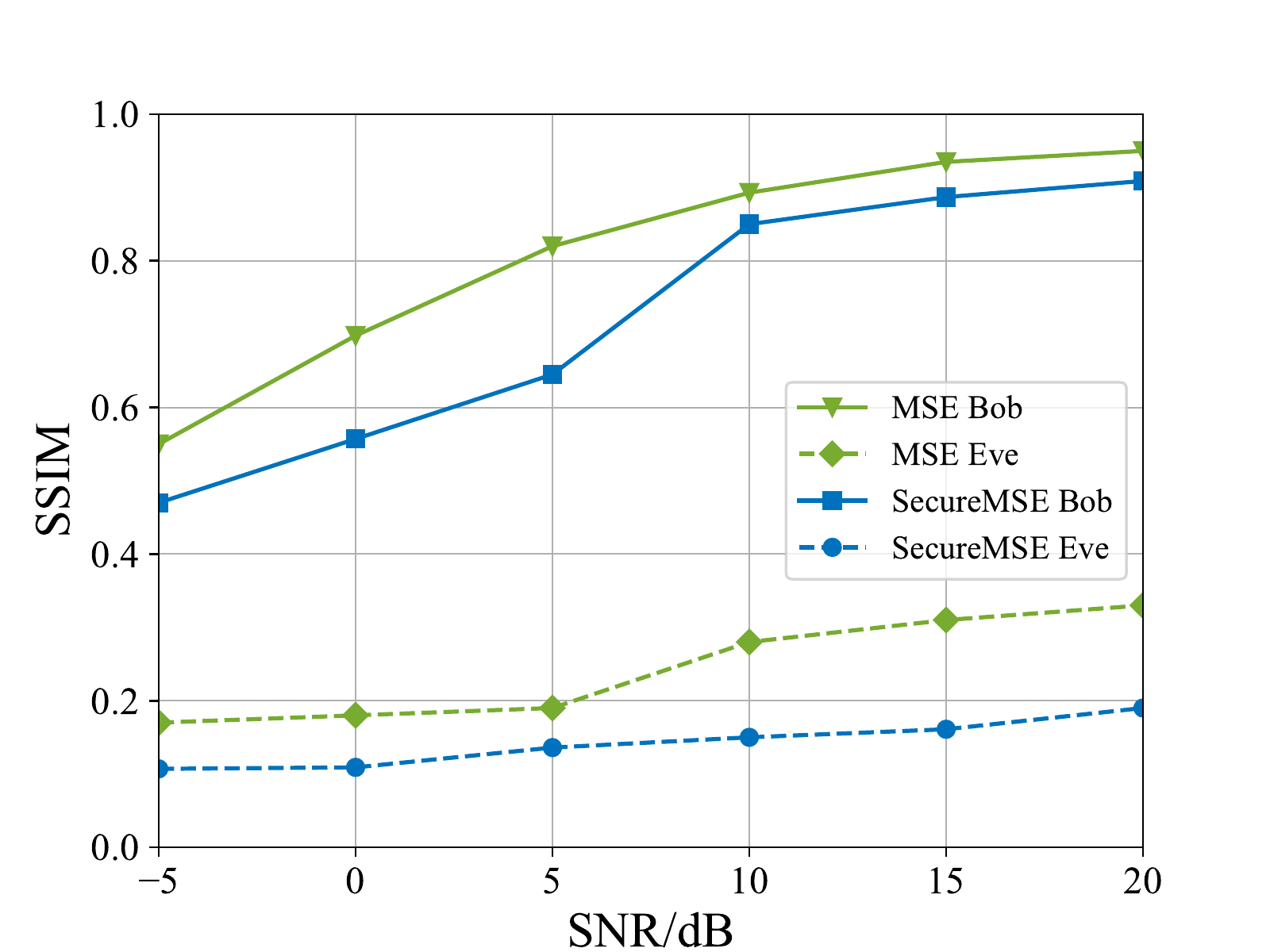}
  \caption{Performance comparison of Bob and Eve in MISO system}\label{fig: secure SSIM MISO}
    %\vspace{-10mm}
  \end{figure}
  \begin{figure*}
    \centering
    \subfigure[Original image]{
      \begin{minipage}[t]{0.19\linewidth}
        \centering
        \includegraphics[width=1\linewidth]{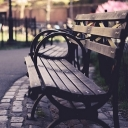}\\
        \vspace{0.02cm}
        % \includegraphics[width=1\linewidth]{Figures/Base_AWGN/d2dSNR_10/origin_9.png}\\
        % \vspace{0.02cm}
      \end{minipage}%
    }%
    \subfigure[MSE Bob]{
      \begin{minipage}[t]{0.19\linewidth}
        \centering
        \includegraphics[width=1\linewidth]{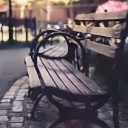}\\
        \vspace{0.02cm}
        % \includegraphics[width=1\linewidth]{Figures/Base_AWGN/d2dSNR_10/predict_9.png}\\
        % \vspace{0.02cm}
      \end{minipage}%
    }%
    \subfigure[MSE Eve]{
      \begin{minipage}[t]{0.19\linewidth}
        \centering
        \includegraphics[width=1\linewidth]{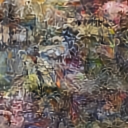}\\
        \vspace{0.02cm}
        % \includegraphics[width=1\linewidth]{Figures/Base_AWGN/d2eavSNR_10/predict_9.png}\\
        % \vspace{0.02cm}
      \end{minipage}%
    }%
    \subfigure[SecureMSE Bob]{
      \begin{minipage}[t]{0.19\linewidth}
        \centering
        \includegraphics[width=1\linewidth]{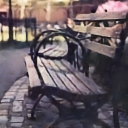}\\
        \vspace{0.02cm}
        % \includegraphics[width=1\linewidth]{Figures/Secure_AWGN/d2dSNR_10/predict_9.png}\\
        % \vspace{0.02cm}
      \end{minipage}%
    }%
    \subfigure[SecureMSE Eve]{
      \begin{minipage}[t]{0.19\linewidth}
        \centering
        \includegraphics[width=1\linewidth]{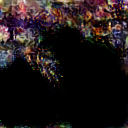}\\
        \vspace{0.02cm}
        % \includegraphics[width=1\linewidth]{Figures/Secure_AWGN/d2eavSNR_10/predict_9.png}\\
        % \vspace{0.02cm}
      \end{minipage}%
    }%
    \centering
    \caption{Examples of reconstructed images produced by Bob and Eve targeting different objective in AWGN system (${\rm SNR}=10{\rm dB}$)}\label{fig:example of AWGN}
    \vspace{-0.2cm}
\end{figure*}
\vspace{-1mm}
\subsubsection{AWGN System}Under the AWGN system, the reconstruction performance of the model based on the two training objectives (e.g., MSE in (\ref{eq:mse loss function}), SecureMSE in (\ref{eq: secure mse loss definition})) is shown in Fig. \ref{fig: secure SSIM AWGN}. 
It is found that the MSE scheme achieves the best reconstruction perforamance at Bob side. 
However, with the increase of SNR, especially when ${\rm SNR\geq10dB}$, the reconstruction performance of Eve improves a lot as well. 
We present some examples of reconstructed images when ${\rm SNR=10dB}$, it can be seen that baseline models without privacy awareness can also roughly reconstruct the approximate image, thus verifying the existence of privacy leakage in the current semantic communication system. 
For the proposed SecureMSE scheme, the reconstruction quality of Eve does not improve a lot as the SNR grows due to the privacy awareness embedded in the well-designed loss function.  
%effect of Eve become worse a lot since the reconstruction performance of Eve is also considered  in the training objective. 
From Fig. \ref{fig:example of AWGN}, it can be seen that Eve with SecureMSE model can no longer obtain any privacy information. 
In addition, comparing the reconstruction effect of Bob under two objectives, SecureMSE only causes a slight performance loss, which is negligible for human’s perception. 
The validity of the proposed algorithm is thus verified.

\begin{figure*}[!]
  \centering
  \subfigure[Original image]{
    \begin{minipage}[t]{0.19\linewidth}
      \centering
      \includegraphics[width=1\linewidth]{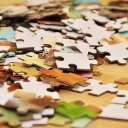}\\
      \vspace{0.02cm}
      % \includegraphics[width=1\linewidth]{Figures/Base_MisoFading/d2dSNR_10/origin_5.png}\\
      % \vspace{0.02cm}
    \end{minipage}%
  }%
  \subfigure[MSE Bob]{
    \begin{minipage}[t]{0.19\linewidth}
      \centering
      \includegraphics[width=1\linewidth]{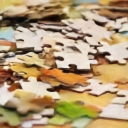}\\
      \vspace{0.02cm}
      % \includegraphics[width=1\linewidth]{Figures/Base_MisoFading/d2dSNR_10/predict_5.png}\\
      % \vspace{0.02cm}
    \end{minipage}%
  }%
  \subfigure[MSE Eve]{
    \begin{minipage}[t]{0.19\linewidth}
      \centering
      \includegraphics[width=1\linewidth]{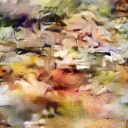}\\
      \vspace{0.02cm}
      % \includegraphics[width=1\linewidth]{Figures/Base_MisoFading/d2eavSNR_10/predict_5.png}\\
      % \vspace{0.02cm}
    \end{minipage}%
  }%
  \subfigure[SecureMSE Bob]{
    \begin{minipage}[t]{0.19\linewidth}
      \centering
      \includegraphics[width=1\linewidth]{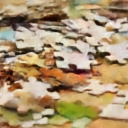}\\
      \vspace{0.02cm}
      % \includegraphics[width=1\linewidth]{Figures/Secure_MisoFading/d2dSNR_10/predict_5.png}\\
      % \vspace{0.02cm}
    \end{minipage}%
  }%
  \subfigure[SecureMSE Eve]{
    \begin{minipage}[t]{0.19\linewidth}
      \centering
      \includegraphics[width=1\linewidth]{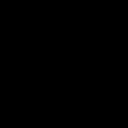}\\
      \vspace{0.02cm}
      % \includegraphics[width=1\linewidth]{Figures/Secure_MisoFading/d2eavSNR_10/predict_5.png}\\
      % \vspace{0.02cm}
    \end{minipage}%
  }%
  \centering
  \caption{Examples of reconstructed images produced by Bob and Eve targeting different objective in MISO system (${\rm SNR}=10{\rm dB}$)}\label{fig:example of MISO}
  \vspace{-0.2cm}
\end{figure*}
\subsubsection{MISO System}
For a typical MISO system, $\mathbf{y}_b$ and $\mathbf{y}_e$ are respectively given by 
\begin{align}
  \mathbf{y}_b = \left(\mathbf{h}_b^H \mathbf{v}\right) \otimes \mathbf{x} + \mathbf{n}_b,~~\mathbf{y}_e = \left(\mathbf{h}_e^H \mathbf{v}\right) \otimes \mathbf{x} + \mathbf{n}_e,
\end{align} 
where $\mathbf{h}_{b}\in \mathbb{C}^{N \times 1}$ and $\mathbf{h}_e\in \mathbb{C}^{N\times 1}$ denote the channel between BS and Bob, BS and Eve, respectively.  
The \emph{maximum ratio transmission} (MRT) precoding scheme is adopted, that is, $\mathbf{v}=\frac{\mathbf{h}_b}{\left\|\mathbf{h}_b\right\|_2^2}$. 
Then, $\mathbf{y}_b$ and $\mathbf{y}_e$ can be rewritten as 
\begin{align}\label{eq: results after precoding }
  \mathbf{y}_b = \mathbf{x} + \mathbf{n}_b,\qquad          \mathbf{y}_e = \alpha_e \mathbf{x} + \mathbf{n}_e, 
\end{align}
where $\alpha_e = \frac{\mathbf{h}_e\mathbf{h}_b^H}{\left\|\mathbf{h}_b\right\|_2^2}$. 
In the following experiment, we has $N=8$, and $P=\frac{\sigma_e^2}{\sigma_b^2}=1$. 

The performance comparison between the proposed scheme and the baseline one under MISO system is shown in Fig. \ref{fig: secure SSIM MISO}. 
In the low SNR regime (i.e., ${\rm -5dB<SNR<5dB}$), the reconstruction performance of SecureMSE in Bob decreases to a certain extent. 
{This is because in the MISO system, as shown in (\ref{eq: results after precoding }), the main difference between Bob and Eve is the fading coefficient. 
}
With the high noise level, the JSC decoder can not distinguish Bob and Eve from the heterogeneity of channel, 
thus makes it more difficult to maximize the reconstruction quality at Bob while suppressing the privacy leakage to Eve. 
As the SNR increases (i.e., ${\rm SNR>=10dB}$), it can be seen that both the reconstruction performance of Bob and the privacy preservation against Eve improve a lot. 
In addition,  the example  reconstruction image with $\rm SNR = 10dB$ is shown in Fig. \ref{fig:example of MISO}. It can be seen that the model based on the conventional MSE loss still suffer from the problem of privacy leakage, while the model trained with the proposed  SecureMSE loss once again prevents privacy leakage, thus verifying the robustness of the proposed algorithm in dealing with various scenarios.
\vspace{-3mm}      
\section{Conclusion}
In this letter, we study the semantic communication system for wireless image transmission, and an efficient JSC framework is developed. 
In addition, we discuss the privacy issue in the current semantic communication system and reveal the potential privacy leakage. % has privacy leakage problems in many scenarios. 
Prompted by this, we proposed a data-driven privacy protection scheme called SecureMSE featuring a well designed loss function with privacy awareness. 
Experimental results verify the effectiveness and robustness of the proposed scheme.
\bibliographystyle{ieeetr}
\vspace{-2mm}
\bibliography{BibDesk_File_v2}
\vspace{-2mm}

\end{document}